\documentclass[aps, reprint, pra, nofootinbib, superscriptaddress]{revtex4-2}

\usepackage[utf8]{inputenc}
\usepackage{float}
\usepackage{graphicx}  
\usepackage{physics}
\usepackage{esint}
\usepackage{amssymb}  
\usepackage{mathtools}
\usepackage{amsmath}
\usepackage{amsthm}
\usepackage{mathrsfs}
\usepackage{bm}
\usepackage[normalem]{ulem}

\usepackage[dvipsnames]{color,xcolor,colortbl}
\usepackage{hyperref}
\usepackage{url}
\hypersetup{%
    colorlinks=true,
    linkcolor=NavyBlue,
    citecolor=NavyBlue,
    urlcolor=NavyBlue,
    plainpages=false
    }

\usepackage[most]{tcolorbox}  

\newtcolorbox[auto counter]{thmbox}[1]{colback=white,
colback=White!99!Black,
colframe=Black!100!black,fonttitle=\bfseries,
title={#1}, enhanced}

\newcommand{\edit}[1]{{\color{black}#1}}

\begin{document}


\title{Physical-Layer Machine Learning with Multimode Interferometric Photon Counting}

\author{Jia-Jin Feng}

\author{Anthony J. Brady}
\affiliation{
Ming Hsieh Department of Electrical and Computer Engineering, University of Southern California, Los
Angeles, California 90089, USA
}

\author{Quntao Zhuang}\email{qzhuang@usc.edu}
\affiliation{
Ming Hsieh Department of Electrical and Computer Engineering, University of Southern California, Los
Angeles, California 90089, USA
}
\affiliation{Department of Physics and Astronomy, University of Southern California, Los Angeles, California 90089, USA}


\begin{abstract}
The learning of the physical world relies on sensing and data post-processing. When the signals are weak, multidimensional and correlated, the performance of learning is often bottlenecked by the quality of sensors, calling for integrating quantum sensing into the learning of such physical-layer data. An example of such a learning scenario is the stochastic quadrature displacements of electromagnetic fields, modeling optomechanical force sensing, radiofrequency photonic sensing, microwave cavity weak signal sensing, and other applications. We propose a unified protocol that combines machine learning with interferometric photon counting to reduce noise and reveal correlations. By applying variational quantum learning with multimode programmable quantum measurements, we enhance signal extraction. Our results show that multimode interferometric photon counting outperforms conventional homodyne detection proposed in prior works for tasks like principal component analysis (PCA) and cross-correlation analysis (CCA), even below vacuum noise levels. To further enhance the performance, we also integrate entanglement-enhanced modules, in the form of squeezed state distribution and anti-squeezing at detection, into the protocol. Combining multimode interferometric photon counting and multipartite entanglement, the proposed protocol provides a powerful toolbox for learning weak signals.
\end{abstract}

\date{\today}

\maketitle



\section{Introduction}

The learning of the physical world involves making measurements to gather data and post-processing the data to obtain information or distill laws of physics. When the physical signals under study are ``strong", such that data from the measurements are precise, then the major task of learning involves classical data processing. In such cases, classical machine learning algorithms, such as support-vector machine (SVM), principle component analysis (PCA) and neural-network based algorithms, often resolve such problems. Various quantum machine learning works~\cite{rebentrost2014quantum,lloyd2014quantum,biamonte2017quantum} have focused on adopting quantum algorithms to assist the processing of classical data. 

In contrast, when the physical signals are ``weak", the quality of the data fundamentally limits the performance of the entire learning process, regardless of sophisticated classical or quantum post-processing algorithms. In this regime, quantum sensing~\cite{giovannetti2011advances,degen2017,pirandola2018advances,Zhang2021DQSrvw} becomes central to enhancing the data acquisition precision, speed or bandwidth, through the use of quantum resources such as squeezing, entanglement, or nonclassicality. However, improving data acquisition can be resource-intensive for large data sets, especially when only low-dimensional features of the dataset are ultimately needed. In this regard, integrating the quantum sensing and post-processing into a single module provides a resource-efficient quantum advantage in learning the physical world, as exemplified by the frameworks of Supervised Learning Assisted by an Entangled sensor Network (SLAEN)~\cite{Zhuang2019SLAEN_theory,Xia2021PRX_SLAEN,liao2024quantum} and quantum computational sensing~\cite{allen2025quantum, Saeed2025qcs, Saeed2025ad}.

\begin{figure*}
    \centering
    \includegraphics[width=.95\linewidth]{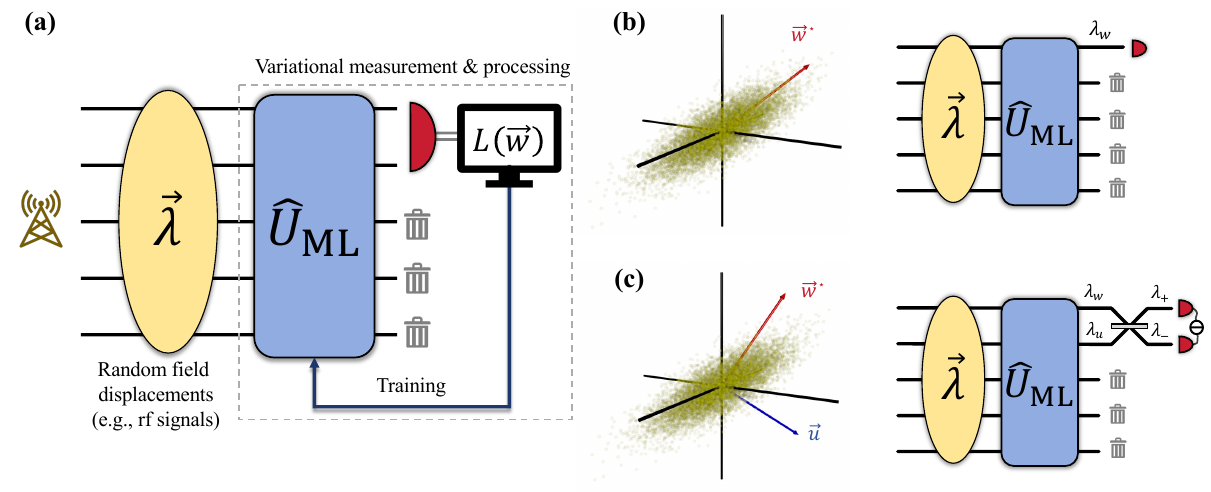}
    \caption{Variational quantum processing of stochastic displacement data. (a) High-dimensional random displacement data --- e.g., from radio-frequency (rf) signals~\cite{Xia202PRL_RadioQSN} or opto-mechanical signals~\cite{Xia2023OmechDQS} --- are collected, measured, and processed to extract certain properties from the data, such as principal components and cross-correlations, using a variational measurement and processing unit. This unit comprises of a programmable linear optical circuit ($\hat{U}_{\rm ML}$), photon counters, and a classical post-processing and machine learning module. Illustrative data (yellow dots) and corresponding measurement configurations are shown for (b) Principal Component Analysis (PCA) and (c) Collective Cross-correlation Analysis (CCA).}
    \label{fig:main}
\end{figure*}

A representative example is that of small stochastic quadrature displacements of a resonator (e.g., mechanical oscillator or optical mode) with amplitudes below the vacuum fluctuations. Such signals appear in precision force sensing~\cite{Xia2023OmechDQS}, gravitational wave sensing~\cite{LIGO2019PRL_QuEnhancedLIGO,Gardner2024PRL_Waveform}, radiofrequency (RF) detection~\cite{Xia202PRL_RadioQSN}, and dark matter searches~\cite{HAYSTAC2021QuDMSearch,Brady2022QuNetworkDMSearch,Shi2023DMLimits}, and have been increasingly studied in distributed quantum sensing~\cite{Brady2024CorrNoise}. While many learning strategies in this context adopt squeezing and homodyne detection~\cite{Zhuang2019SLAEN_theory,Xia2021PRX_SLAEN} (and related learning works~\cite{Oh2024LearnOscillators,Liu2025bosonicLearning}) --- often inspired by coherent displacement sensing protocols --- recent results suggest that these tools are strictly suboptimal for weak stochastic signals. 
This motivates the framework of quantum computational sensing, which leverages sequences of quantum sensing, quantum operations, measurements, and optimization to achieve both reduced hardware overhead and lower error probability~\cite{Saeed2025ad,Saeed2025qcs}.
In particular, photon counting measurements are known to provide a quantum advantage in the single-mode case~\cite{Shi2023DMLimits, Dixit2021QubitDMSearch}, but their role in high-dimensional, variationally trained architectures remains largely unexplored.

In this paper, we introduce a framework for physical-layer supervised learning of weak stochastic displacement signals using \textit{multimode interferometric photon counting} (Fig.~\ref{fig:main}). Our architecture combines a programmable linear optical (a.k.a., beamsplitter) network and photon counting detectors which are optimized via variational training to solve various learning tasks. \edit{Our module thus serves a variational quantum measurement toolbox~\cite{Meyer2021:varToolbox}.} As examples, we consider the tasks of PCA and Cross-Correlation Analysis (CCA), which are relevant to data compression and correlation estimation of high-dimensional data \edit{\cite{Masiero2009}. These machine-learning techniques are widely used in experiments, for instance, in image recognition tasks implemented on a nuclear magnetic resonance (NMR) quantum processor \cite{Xin2021, Dominik2024}. Our framework applies to some of these application scenarios in force and field sensing with mechanical sensors as well as thermal-field imaging, as we discuss further in Section~\ref{sec:phys-relev}.}

Beyond the advantage attained via interferometric photon counting, we explore enhancements from multipartite entanglement, realized by distributed squeezed-vacuum states (Fig.~\ref{fig:squeezed}, see also Refs.~\cite{Zhuang2018PRA_dqs,Zhang2021DQSrvw}). The resulting \textit{squeezed-vacua-assisted variational processor} features a sandwich structure, with programmable linear-optical networks enclosed between (single-mode) squeezers and anti-squeezers. This architecture coherently amplifies weak displacement signals while preserving the multimode interference established by the linear-optical network, enabling a further enhancement in learning tasks. The result is a robust and scalable physical-layer learning protocol that combines entanglement-enhanced sensing with task-oriented optimization.


\section{Problem set up}

In this section, we provide a high-level overview of the setup and a description of the quantum data acquisition and processing framework. We spell out the problems of interest in the guise of variational (i.e., adaptive) quantum parameter estimation and delineate two detection strategies, homodyne and photon counting, for comparison.


Consider that $N$ quantum probes (e.g., optical modes) traverse a quantum channel which encodes valuable quantum data into the probe states in the form of stochastic quadrature displacements. Physically, signals from RF-photonics, optomechanics, complex microwave cicuit networks, or other sources may constitute the data. To process the outgoing state encoded with quantum data, we apply a tunable, multimode variational quantum circuit. After which, measurements are performed on the processed state to construct estimates of the encoded data. This process iterates adaptively: the variational circuit parameters are tuned based on the estimated data to optimize downstream tasks. For instance, by refining the circuit, we aim to enhance the extraction of principal components (principle component analysis, PCA) or high-dimensional cross correlations directly at the physical layer. Below, we formalize the setup.

\subsection{Physical-layer quantum data} 

We assume that the quantum data takes the form of random momentum displacements, denoted by the vector $\vec{\lambda}$, within an $M$-mode quantum system. We model the displacements by a zero-mode random displacement quantum channel, $\Phi_{\bm{V}}$, characterized by the covariance matrix $\bm{V}_{ij} = \mathbb{E}[\vec{\lambda}_i \vec{\lambda}_j]$ (and $\mathbb{E}[\vec{\lambda}]=0$).\footnote{The signal can follow a general distribution; however, we focus on signals characterized entirely by their covariance matrix. This assumption simplifies both analytical and numerical analysis. } Motivated by optomechanical sensing, we assume all displacements occur along $M$ momentum quadratures, though one may extend our framework to include conjugate position quadratures. The momentum operator vector is given by $\hat{\vec{P}}\equiv i\left( \hat{\vec{a}}^\dagger-\hat{\vec{a}} \right)/\sqrt{2}$, where $\hat{\vec{a}}$ is the photon annihilation operator for the $M$ channels. The channel acts instance-wise as $\Phi_{\bm{V}}: \hat{\vec{P}} \rightarrow \hat{\vec{P}} + \vec{\lambda}$, where $\vec{\lambda}$ is the stochastic displacement with $\mathbb{E}[\lambda_i \lambda_j] = \bm{V}_{ij}$. Typically, the quantum data are high-dimensional, encoded across many modes ($M \gg 1$). Furthermore, we assume operation in the quantum regime (i.e., low signal-to-noise ratio), where \(|\bm{V}| \ll 1\); that is, the typical magnitude of momentum fluctuations is significantly smaller than vacuum fluctuations, which are of order 1.


\edit{This framework can be naturally extended to handle simple isotropic displacements affecting both quadratures equally, i.e., identical shifts in $Q$ and $P$, which occurs in RF-photonics~\cite{Xia202PRL_RadioQSN, Shi2025:RFimaging}, optical imaging~\cite{Tsang2016Superresolution, Tsang2019Starlight, Lupo2020PRL_QuLinearLimits, Grace2022Imaging, Buonaiuto2025ML_subdiff}, and microwave cavity settings~\cite{HAYSTAC2021QuDMSearch,Brady2022QuNetworkDMSearch,Agrawal2024FockRadiometer}. In this case, the analysis proceeds analogously to the single-quadrature scenario, with the relevant covariance matrix reflecting the combined displacements. Asymmetric displacements which differ between $Q$ and $P$ are more challenging to address, as photon counting is insensitive to this asymmetry and, thus, separating quadrature-specific fluctuations might require adaptive measurements or additional probe resources.}

\subsection{Variational quantum processing}
To begin, we focus on the ``passive" sensing scenarios. That is, we vary the detection (receiver) while considering the probes being in their vacuum state. This is motivated by weak stochastic signal sensing, where vacuum and photon counting provides a significant advantage over quantum sensing schemes based on homodyne or heterodyne detection~\cite{Shi2023DMLimits,Dixit2021QubitDMSearch}. In Section~\ref{sec:squeezing}, we extend towards further quantum advantage from multipartite entanglement via distributed squeezing.

The quantum channel $\Phi_{\bm{V}}$ induces random displacements $\vec{\lambda}$ on the vacuum probes, which are subsequently processed by the linear optical variational circuit $\hat{U}_{\rm{ml}}$ (see Fig.~\ref{fig:main}). The role of the variational quantum processor $\hat{U}_{\rm{ml}}$ is to coherently combine the stochastic displacement amplitudes before measurement. Mathematically, we characterize the variational circuit by the orthogonal matrix \(\bm{B} = (\vec{w}, \vec{u}, \dots)^T\), where \(\vec{w}\) and \(\vec{u}\) are orthonormal vectors defining the transformation (additional vectors in $\bm{B}$ may be included as needed). When acting on vacuum probes, the processor effectively implements the operator transformation $\hat{\vec{P}} \rightarrow \hat{\vec{P}} + \bm{B} \vec{\lambda}$, which converts the original displacement channel $\Phi_{\bm{V}}$ into the new multimode displacement channel $\Phi_{\bm{B} \bm{V} \bm{B}^T}$. At the physical layer, $\vec{w}$ and $\vec{u}$ correspond to collective displacements, $\lambda_w=\vec{w}^T\vec{\lambda}$ and $\lambda_u=\vec{u}^T\vec{\lambda}$, that carry information about the non-local fluctuations and correlations within the system.

Detectors measure the output probes, yielding estimates of the collective displacement data. These estimates inform the updates to the variational circuit (i.e., the parameters of $\bm{B}$) to minimize a loss function, and thus optimize data processing for tasks like PCA and CCA (see below). This iterative feedback thus trains and refines the variational circuit based on sensor-derived data.  

In conventional machine learning, the loss function is well-defined, and gradient-based optimization is often effective. However, in quantum systems, measurement-induced uncertainty introduces substantial quantum noise into the loss function, especially in the weak signal case of interest. This noise severely hampers gradient-based optimization, as small parameter updates can lead to disproportionately large fluctuations in the estimated gradient. To address this challenge, our protocol incorporates this inherent poisoning effect by employing a gradient-free optimization algorithm. 

We detail the training process in Section~\ref{sec:training}.

\subsection{Detection Strategies.}
The performance of the variational circuit for various tasks depends crucially on the measurement schemes that we employ. We consider two forms of measurement for comparison: homodyne and photon counting.

The standard for measuring displacements and quadrature fluctuations is homodyne detection. In this case, the output amplitudes of the variational circuit are directly measured. Notably, this implies that we may perform local homodyne measurements on all modes and then combine the measurement results in classical post-processing, viz., constructing bilinear estimators from the measurement data to learn properties of $\bm V$. Consequently, the unitary $\hat{U}_{\rm ML}$ need not be physically implemented, as its action can be emulated in post-processing. Nevertheless, we retain $\hat{U}_{\rm ML}$ in all figures to maintain consistency across measurement protocols.\footnote{The physical implementation of $\hat{U}_{\rm ML}$ remains relevant in scenarios where, for instance, the number of available homodyne detectors is limited or measurement efficiency is a limiting factor.}

Photon counting corresponds to intensity detection, i.e. measuring squared amplitudes. Thus processing via $\hat{U}_{\rm ML}$ prior to (local) photon counting measurements is absolutely essential --- otherwise non-trivial joint measurements are required. Observe that photon counting is inherently a non-Gaussian measurement. Therefore, when applied after the variational circuit $\hat{U}_{\rm ML}$, the total measurement scheme effectively realizes a complex, non-Gaussian measurement, which we refer to as multimode interferometric photon counting; see Fig.~\ref{fig:main} for reference. 

\edit{We focus on homodyne detection and photon counting in this work because they are experimentally accessible and widely implemented. Importantly, a variational linear-optical circuit followed by photon-counting detection already constitutes a type of joint non-Gaussian measurement, as local photon counting alone cannot achieve the same performance. Other joint non-Gaussian measurements are possible but are generally more complex and challenging to realize.
\\
\indent
Moreover, this measurement scheme is particularly relevant for the tasks studied here, as it is known to be near-optimal for estimating stochastic displacements~\cite{Gorecki2022SpreadChannel, Shi2023DMLimits} and for imaging weak thermal fields~\cite{Lupo2020PRL_QuLinearLimits}. Heterodyne detection could serve as a useful benchmark as well, but the expected performance is on par with homodyne due to the inherent linearity of the measurement.}

\subsection{Interesting Problems (PCA and CCA)}
In this work, we focus on two specific problems. The first is Principal Component Analysis (PCA)~\cite{Bailey_2012}, where the objective is to identify the principal component $\vec{w}^\star$ of the covariance matrix $\bm{V}$ such that $\vec{w}^\star = \arg\max_{\vec{w}} (\vec{w}^T \bm{V} \vec{w})$. Hence, $\vec{w}^\star \bm{V}\vec{w}^\star$ represents the maximal variance of $\bm V$. The second problem is what we call \textit{Collective Cross-correlation Analysis} (CCA), in which, given a fixed reference vector $\vec{u}$, we seek the orthonormal vector $\vec{w}^\star$ that maximizes the cross-correlation magnitude, i.e., $\vec{w}^\star={\rm argmax}_{\vec{w}}(|\vec{u}^T\bm{V}\vec{w}|)$. In both problems, the converged quantum circuit set-up provides vector information on component of interest, while the resulting measurement provides additional amplitude information. Formally, we can define the PCA and CCA variables for a fixed variational circuit as
\begin{eqnarray}
    \vartheta_{\rm PCA}(\vec{w})\coloneqq\vec{w}^T \bm{V} \vec{w} \qq{and} \vartheta_{\rm CCA}(\vec{w})\coloneqq\vec{u}^T \bm{V} \vec{w},
\end{eqnarray}
which represent the variance and cross correlation between the collective stochastic displacements $\lambda_w$ and $\lambda_u$, respectively. 

Our analyses thus aim to identify prominent statistical structures in high-dimensional correlated data and reveal meaningful properties thereof. In the next section, we describe how such information can be accessed through physical-layer measurement protocols with high precision, even when the underlying physical signals are weak.

\edit{\subsubsection{Physical Relevance}}
\label{sec:phys-relev}

\edit{
Our motivation for focusing on PCA and CCA is that, in many quantum sensing settings, data processing is inseparable from the physical measurement itself: The most informative physical quantities of interest (the signals) often appear in collective or correlated modes, rather than in the raw measurement basis. From this perspective, PCA and CCA are not merely pre-processing tools but form part of the measurement strategy, identifying which physical modes to access and how to combine them. Below we point out representative scenarios where our analysis may be useful.
\\
\paragraph*{\textbf{Mechanical field/force sensing.}} In arrays of trapped ions~\cite{Gilmore2021IonEFieldQSN} or optomechanics~\cite{Xia202PRL_RadioQSN, Brady2023OmechArray, Xia2023OmechDQS}, the raw measurement data naturally lives in the position basis of individual oscillators. However, the physically relevant signals, originating from background fluctuating fields or forces, often manifest most strongly in a small number of collective modes, such as the center-of-mass or antisymmetric modes. PCA provides a natural way to extract these dominant modes, while our CCA formulation illustrates how correlations between distinct collective modes can be probed within the same variational framework.
\\
\paragraph*{\textbf{Thermal-field imaging.}} Detection scenarios in which the collected field is stochastic and obeys thermal photon statistics arise in many physical settings. Examples include incoherent fluorescence imaging~\cite{Tsang2016Superresolution, Tsang2019Starlight, Lupo2020PRL_QuLinearLimits, Grace2022Imaging, Buonaiuto2025ML_subdiff}, sensing with a microwave cavity network~\cite{Brady2022QuNetworkDMSearch}, and radio-frequency imaging~\cite{Xia202PRL_RadioQSN, Shi2025:RFimaging}. In these multimode settings, the raw basis rarely provides direct access to the most informative physical quantities of interest, particularly when the signals are small. As known in the quantum imaging literature~\cite{Tsang2019Starlight}, the signals may be concentrated in only a few structured or collective modes (e.g., Hermite-Gaussian). Identifying this basis is therefore crucial, and PCA serves as a natural tool for this task.}

\section{Protocols and Training Methods}
\label{sec:training}

In this section, we present our measurement protocols and training methods for both PCA and CCA, considering homodyne detection and photon counting as alternative measurement strategies. To frame our approach, we begin with a broad qualitative overview. 

Training in our setting is subject to two distinct sources of noise: (i) classical noise from sampling the stochastic displacement signals, and (ii) quantum noise arising from measurements of the output quantum state. Both forms of noise obscure the loss function landscape --- blurring gradient information and masking the true optimum --- which can hinder convergence and thus complicate optimization.

To systematically account for noise, we define the loss function as a quantum observable, denoted by $\hat{L}\coloneqq\hat{L}(\vec{w})$, which depends on the variational circuit via $\vec{w}$. When convenient, we suppress the explicit dependence on the variational circuit for brevity. While the specific form of $\hat{L}$ is determined by the detection strategy and learning task, we generally choose it such that its expectation value satisfies $\expval*{\hat{L}}=-\vartheta_{\rm{PCA/CCA}}$. The training objective is thus to minimize $\expval*{\hat{L}}$ by optimizing the parameters of the variational circuit, thereby estimating the optimal value $\vartheta^\star_{\mathrm{PCA/CCA}}$ from physical measurements of the loss observable. Since $\hat{L}$ is a quantum operator, its measured values inherently fluctuate (i.e., ${\rm Var}(\hat{L})\neq 0$). This statistical uncertainty places fundamental limits on the achievable precision. As a result, the detection strategy plays a crucial role in mitigating noise during training.

\subsection{Principle Component Analyzer (PCA)}

The principal component of a (generally high-dimensional) data set is the projection direction that has the maximum variance. In terms of the covariance matrix $\bm{V}$ of the data set, this corresponds to its largest eigenvalue and represents the direction of maximal variance in the data. Our goal in PCA is to estimate this principal eigenvector, denoted $\vec{w}^\star$, directly at the physical layer using quantum measurements. We recast this task as a variational single-mode parameter estimation problem.

To do so, we introduce a trainable vector $\vec{w}$, which we define as the first row of the variational linear optical transformation $\bm{B}$, such that $\vec{w}_i = \bm{B}_{1i}$. We focus exclusively on the first output mode of the variational circuit [see Fig.~\ref{fig:main}(b)], which corresponds to the collective mode defined by $\vec{w}$. The displacement along this mode is given by $\lambda_w = \vec{w}^T \vec{\lambda}$, with associated variance $\vartheta_{\rm PCA}(\vec{w})=\mathbb{E}[\lambda_w^2]= \vec{w}^T \bm{V} \vec{w}$. Given a fixed detection scheme (e.g., homodyne or photon counting), we define an appropriate loss function observable $\hat{L}$ such that $\expval*{\hat{L}} = -\vartheta_{\rm PCA}(\vec{w})$. We then adaptively train the variational circuit to update $\vec{w}$ and minimize our estimate $\expval*{\hat{L}}$, thereby converging toward the principal component $\vec{w}^\star$.

\paragraph*{\textbf{Photon counting.}} 
In the photon counting implementation, we measure the photon number at the first output of the variational processor—i.e., in the mode defined by $\vec{w}$. The loss function observable is defined as
\begin{equation}
    \hat{L} = -2\hat{N}_w,
\end{equation}
where $\hat{N}_w$ is the photon number operator for mode $w$. The output state in this mode is an anisotropic thermal state with mean energy $\mathbb{E}[\lambda_w^2]/2$ --- thus, $\expval*{\hat{L}} = -\mathbb{E}[\lambda_w^2] = -\vartheta_{\rm PCA}(\vec{w})$,
as desired. The noise fluctuations of the loss function are determined by the variance of the photon number,
\begin{equation}\label{eq:VarCount}
    \mathrm{Var}(\hat{L})=4\mathrm{Var}(\hat{N}_w) \approx 2\mathbb{E}[\lambda_w^2],
\end{equation}
which holds in the weak-signal regime.

\paragraph*{\textbf{Homodyne detection.}} 
For homodyne detection, we measure the field amplitude quadrature $\hat{P}_w$ of the mode defined by $\vec{w}$. Because the signal is encoded in the variance of the displacement, the relevant loss function is the squared amplitude,
\begin{equation}\label{eq:VarHom}
    \hat{L} = -\hat{P}_w^2.
\end{equation}
Again, the expectation value yields $\expval*{\hat{L}} = -\mathbb{E}[\lambda_w^2] = -\vartheta_{\rm PCA}(\vec{w})$, in agreement with the PCA objective. However, unlike photon counting, the variance of the loss function includes a constant contribution from vacuum fluctuations, namely,
\begin{equation}
    \mathrm{Var}(\hat{P}_w^2) = 2\mathrm{Var}(\hat{P}_w)^2 \approx 1/2,
\end{equation}
where $\mathrm{Var}(\hat{P}_w)=\mathbb{E}[\lambda_w^2]+1/2\approx 1/2$ was used.

For weak stochastic signals, homodyne detection suffers from a comparatively high and irreducible noise floor. To make this concrete, we compare variances of the loss functions for both detection strategies [see Eqs.~\eqref{eq:VarCount} and~\eqref{eq:VarHom}] and observe that
\begin{equation}\label{eq:compare}
    \frac{{\rm Var}(\hat{L})\big|_{\rm count}}{{\rm Var}(\hat{L})\big|_{\rm hom}}\approx 4\mathbb{E}[\lambda_w^2]\ll 1.
\end{equation}
Lower fluctuations in the loss function translate directly into more efficient and robust PCA training. Therefore, photon counting offers superior sensitivity to homodyne detection in the weak-signal regime, ultimately ensuring more stable convergence during training (see Section~\ref{sec:results}).

\subsection{Collective Cross-Correlation Analysis (CCA)}

The objective of Collective Cross-Correlation Analysis (CCA) is to identify a direction $\vec{w}^\star$ that is maximally correlated with a fixed reference direction $\vec{u}$. [The fixed reference might correspond to a particularly interesting feature direction in our data.] Specifically, we seek to maximize the cross-correlations between the projected displacements $\lambda_u = \vec{u}^T \vec{\lambda}$ and $\lambda_w = \vec{w}^T \vec{\lambda}$ associated with two mode $\vec{u}$ and $\vec{w}$. 

We implement this task at the physical layer by embedding both $\vec{w}$ and $\vec{u}$ into the variational linear optical transformation $\bm{B}$, such that $\vec{w}_i = \bm{B}_{1i}$ and $\vec{u}_i = \bm{B}_{2i}$. We then focus on the first two output modes of the variational circuit [see Fig.~\ref{fig:main}(c)], which thus experience random displacements $\lambda_w$ and $\lambda_u$ respectively. The relevant output data is encoded in the covariance matrix of the modes $\vec{u}$ and $\vec{w}$, which has elements $\mathbb{E}[\lambda_w^2]$, $\mathbb{E}[\lambda_u^2]$, and $\mathbb{E}[\lambda_w \lambda_u]$. We recast this task as a variational two-mode parameter estimation problem.

The signal of interest is the cross-correlation between the two projected modes $\vartheta_{\rm CCA}(\vec{w}) = \mathbb{E}[\lambda_w \lambda_u] = \vec{w}^T \bm{V} \vec{u}$. To estimate this quantity at the measurement stage, we adopt an interferometric readout strategy: the two output modes $\lambda_w$ and $\lambda_u$ are interfered on a balanced (50:50) beamsplitter, followed by photon counting at both output ports. The post-processed difference in photon counts yields access to the cross-correlation.\footnote{A similar strategy applies to homodyne detection, though in that case, ``interference" can be implemented in classical post-processing.}

As in the PCA case, we define an appropriate loss function observable $\hat{L}$ based on the chosen detection scheme, such that $\expval*{\hat{L}}=\vartheta_{\rm CCA}(\vec{w})$. We then iteratively train the variational circuit to update $\vec{w}$ and minimize our estimate $\expval*{\hat{L}}$, thereby converging toward the optimal direction $\vec{w}^\star$ that maximizes the estimated correlation.

\paragraph*{\textbf{Photon counting.}} 
For CCA with photon counting, we introduce a balanced beamsplitter after the variational processor to interfere the two relevant modes, yielding new effective displacements $\lambda_{\pm} = (\lambda_w \pm \lambda_u)/\sqrt{2}$. We then perform photon counting on both output ports [see Fig.~\ref{fig:main}(c)]. The loss function observable is defined as the photon number difference,
\begin{equation}
    \hat{L} = -\frac{1}{2}(\hat{N}_+ - \hat{N}_-),
\end{equation}
where $\hat{N}_\pm$ denote photon number operators for the output modes of the beamsplitter. The factor of $1/2$ ensures that the expectation value corresponds to the cross-correlation of the displacements, i.e., $\expval*{\hat{L}} = -\mathbb{E}[\lambda_w \lambda_u] = -\vartheta_{\rm CCA}(\vec{w})$,
in accordance with the CCA objective. In the weak-signal regime, the variance of the photon number difference becomes
\begin{equation}
    \mathrm{Var}(\hat{N}_+ - \hat{N}_-) \approx \mathbb{E}[\lambda_w^2] + \mathbb{E}[\lambda_u^2].
\end{equation}

\paragraph*{\textbf{Homodyne detection.}} 
For homodyne detection, we directly measure the field quadratures $\hat{P}_w$ and $\hat{P}_u$ of the two relevant modes and compute their product in post-processing. The corresponding loss function is
\begin{equation}
    \hat{L} = -\hat{P}_w \hat{P}_u,
\end{equation}
with expectation value $\expval*{\hat{L}}=-\mathbb{E}[\lambda_w \lambda_u] = -\vartheta_{\rm CCA}(\vec{w})$, again matching the CCA objective. The variance of this estimator can be evaluated straightforwardly using Wick's theorem,
\begin{equation}
    \mathrm{Var}(\hat{P}_w \hat{P}_u) = \langle \hat{P}_w^2 \rangle \langle \hat{P}_u^2 \rangle - \langle \hat{P}_w \hat{P}_u \rangle^2 \approx 1/4,
\end{equation}
where we used $\langle \hat{P}_i^2 \rangle = \mathbb{E}[\lambda_i^2] + 1/2 \approx 1/2$ in the weak-signal limit. Just as in the PCA case, homodyne detection suffers from an irreducible vacuum contribution that dominates the noise budget when the signal is small. Thus, in the weak-signal regime, photon counting provides greater sensitivity for identifying cross-correlated directions, ultimately ensuring more stable convergence during CCA training (see Section~\ref{sec:results}).

\subsection{Training methods}
\label{ssec:training_methods}

Having defined the measurement protocols for PCA and CCA, we now describe how the circuits are trained for the PCA and CCA tasks. The optimization process must contend with both classical and quantum noise, which obscure the training landscape and render convergence challenging. We therefore turn to robust training procedures tailored to our setup.

To concretely illustrate our methods, we consider the case of maximally correlated noise with the covariance matrix $\bm{V} = 2M \sigma_c^2 \vec{v}\,\vec{v}^T$, where $\vec{v} = (1,\dots,1)/\sqrt{M}$ is a normalized vector and $\sigma_c$ characterizes the fluctuation amplitude per mode. At each measurement step, we sample a displacement vector of the form $\vec{\lambda} = \lambda \vec{v}$, with $\lambda \sim \mathcal{N}(0, \sqrt{2M}\sigma_c)$. A single-shot measurement of the circuit output then yields $L_i$.

\edit{There is an unavoidable contribution from finite sampling noise, which cannot be completely eliminated and leads to significant fluctuations in the loss observable. In conventional gradient descent, the gradient error scales as the noise strength divided by the step size. Consequently, attempting to improve accuracy by decreasing the step size paradoxically amplifies the error. To address this limitation, several alternative optimization strategies have been developed that are more robust to noise. One example is particle swarm optimization (PSO), a gradient-free method known for its robustness in noisy environments \cite{LING2016,KIM2021109088,Sundararaj2019}. Other noise-resilient optimization techniques may also serve as viable alternatives.}

In PSO, the optimization process involves $N_{\rm p}$ parallel particles (or trajectories), each evolving over time via `position' $\vec{w}^{(j)}(t)$ and `momentum' $\vec{d}^{(j)}(t)$. At each epoch $t$, the update rules for particle $j$ are:
\begin{eqnarray}
    \vec{w}^{(j)}(t+1)&=&\vec{w}^{(j)}(t)+\vec{d}^{(j)}(t+1), \nonumber\\
    \vec{d}^{(j)}(t+1)&=&m_a \vec{d}^{(j)}(t)+
    r_1 \left(\vec{w}^{(j)}(t)- \vec{w}_{\rm best}(t)\right) \nonumber\\
    &&+r_2\left( \vec{w}^{(j)}(t)- \vec{w}_{\rm g best}(t)\right),
\end{eqnarray}
where $m_a$ is the inertia coefficient, and $r_1, r_2 \in (0, 0.5)$ are random generated numbers for each update. The term $\vec{w}_{\rm best}(t)=\arg\min_{w\in \{w^{(j)}(t)|1\le j\le N_{\rm p}\}}L[w]$ refers to the best configuration (i.e., with the lowest sampled loss function) found by all particle at time $t$. Similarly, $\vec{w}_{\rm gbest}(t)$ is the best global configuration across all particles and all previous epochs.  If $L[\vec{w}_{\rm gbest}(t)]>L[\vec{w}_{\rm best}(t)]$, then it is updated with $\vec{w}_{\rm gbest}(t)=\vec{w}_{\rm best}(t)$.

To suppress spurious updates caused by statistical fluctuations from quantum measurement noise, we introduce a ``forgetting factor" $g \in [0,1]$. If $\vec{w}_{\rm best}^{(j)}(t)$ does not outperform the current global best, the update is smoothed according to
\begin{equation}
    \vec{w}_{\rm gbest}(t) = (1 - g)\, \vec{w}_{\rm gbest}(t-1) + g\, \vec{w}_{\rm best}^{(j)}(t).
\end{equation}
This modification tempers abrupt changes in the optimization landscape and helps stabilize convergence. In the ideal limit of infinite sampling, this method becomes unnecessary and can be neglected by setting $g=0$.

\edit{In classical machine learning, limited training data often leads to overfitting, which degrades the model’s ability to generalize. In the quantum setting, an additional challenge arises: the loss function cannot be exactly evaluated with a finite number of measurements due to inherent quantum randomness. As a result, an incorrect outcome may occasionally appear with a deceptively low loss, while a correct outcome may yield a spuriously high loss. One simple strategy to mitigate this problem is the introduction of “forgetting” in the learning process~\cite{Alyssa2024,Amartya2024}. By reducing the influence of outliers that disproportionately disrupt optimization, forgetting ensures continuous parameter updates and improves the average performance of the model. }

The initial positions $\vec{w}^{(j)}(0)$ are randomly initialized, with all initial displacements $\vec{d}^{(j)}(0)$ set to zero. The first global best position is defined as $\vec{w}_{\rm gbest}(0) = \vec{w}_{\rm best}(0)$. Throughout the optimization process, we take $\vec{w}_{\rm gbest}^{(j)}(t)$ as the final output instead of $\vec{w}_{\rm best}^{(j)}(t)$.


\begin{figure}
\centering
\includegraphics[clip = true, width =\linewidth]{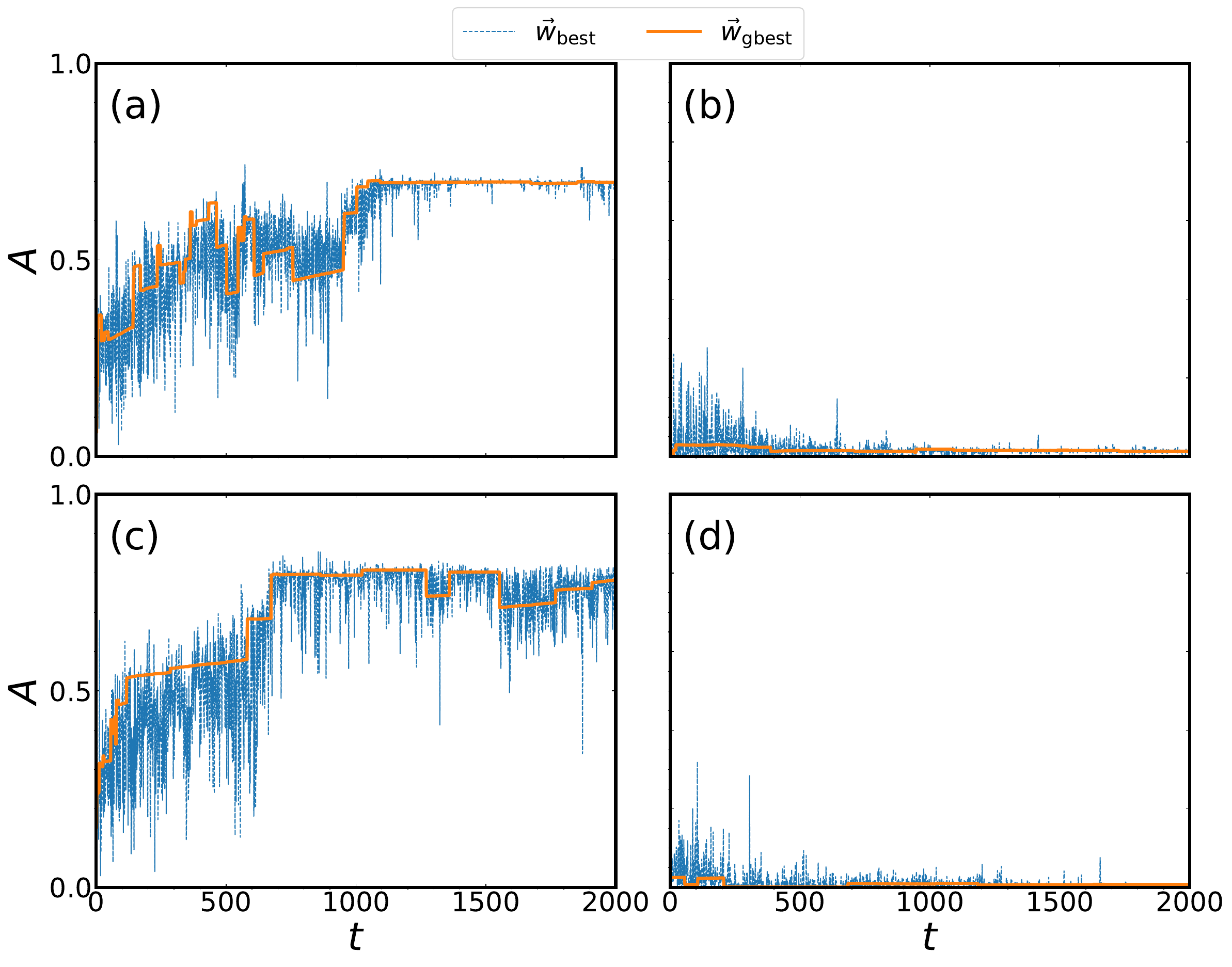}
\caption{\label{fig:Acc}  Training history of the accuracy of (a) PCA with photon counting, (b) PCA with homodyne detection, (c) CCA with photon counting, and (d) CCA with homodyne detection. Other parameters are $M=21$ and $\sigma_c=0.02$. The blue solid curve indicates the accuracy of the best-performing setup at each time step $t$, while the orange dashed curve represents the highest accuracy achieved by the global best configuration up to time $t$.}
\end{figure}

\section{Results and Comparison}
\label{sec:results}

We compare the performance of photon counting and homodyne detection by evaluating the accuracy, defined as the overlap between the optimal vector and the optimized vector,
\begin{equation}        A=\abs{\vec{w}^T\vec{w}^\star}^2.
\end{equation}
This metric provides a direct measure of how closely the variational circuit approaches the true solution. In contrast to the loss function, which can fluctuate due to finite-shot noise, the accuracy provides a better indication of convergence to the ground truth. As we now elaborate, photon counting consistently yields higher accuracy and more robust training than homodyne detection for both PCA and CCA tasks. To begin, we focus on a fixed value of signal strength $\sigma_c=0.02$ and number of modes $M=21$ and present the training history of the accuracy versus time steps $t$ in Fig.~\ref{fig:Acc}. In subplots (a)(c), we see a substantial increase of the accuracy as training proceeds in the case of photon counting, despite fluctuations coming from noises in loss function estimation. In contrast, (b)(d) shows a poor convergence of the training for homodyne detection. \edit{Note that even for the cases with photon counting in subplots (a)(c), the convergence towards unity accuracy is ultimately impeded by the measurement noise from finite sampling.} 

\begin{figure}
    \centering
    \includegraphics[clip = true, width =1\linewidth]{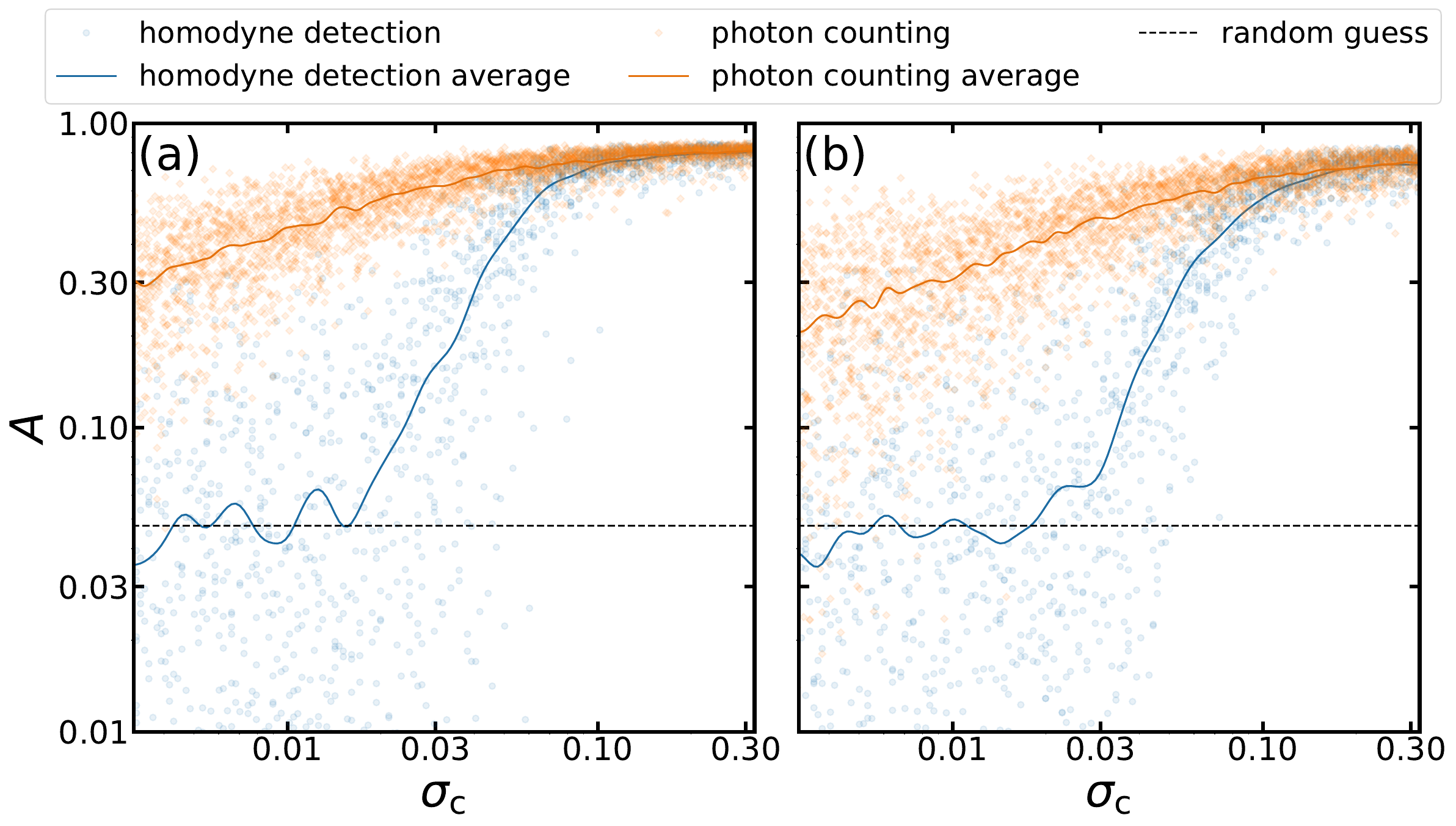}
    \caption{\label{fig:CN} Accuracy depending on signal strength. Each data point corresponds to a different initial condition for optimization and distinct sampling results. The number of modes is $M=21$. The two cases are (a) PCA (b) CCA.}
\end{figure}

To examine performance in the weak-signal regime more thoroughly, we vary the stochastic signal amplitude $\sigma_c$ and plot the resulting accuracy in Fig.~\ref{fig:CN}. As expected, photon counting (orange) significantly outperforms homodyne detection (blue) when the signal amplitude is small ($\sigma_c\ll 1$). While homodyne detection is practically simpler --- relying mainly classical post-processing of quadrature (a.k.a., amplitude) measurements --- it is fundamentally limited by vacuum fluctuations, which dominate the noise budget when $\sigma_c \ll 1$. Contrariwise, photon counting is not subject to vacuum noise: The vacuum state yields zero photons, and a weak stochastic displacement has a small probability ($\sigma_c \ll 1$) of adding one photon to the system of modes, which is directly resolvable by photon counting.

Homodyne accuracy degrades rapidly as $\sqrt{M}\sigma_c\ll 1$, a threshold below which vacuum fluctuations overwhelm the total signal. Photon counting maintains good performance in this regime, degrading more gradually with decreasing signal strength. Only in the limit $\sigma_c \rightarrow 0$ (which is beyond the scope of the plot) do both methods converge to the same baseline: The expected squared overlap between a random unit vector and the target vector, which scales as $1/M$ and is referred to here as the \textit{random guess limit}.

\begin{figure}
    \centering
    \includegraphics[clip = true, width =1\linewidth]{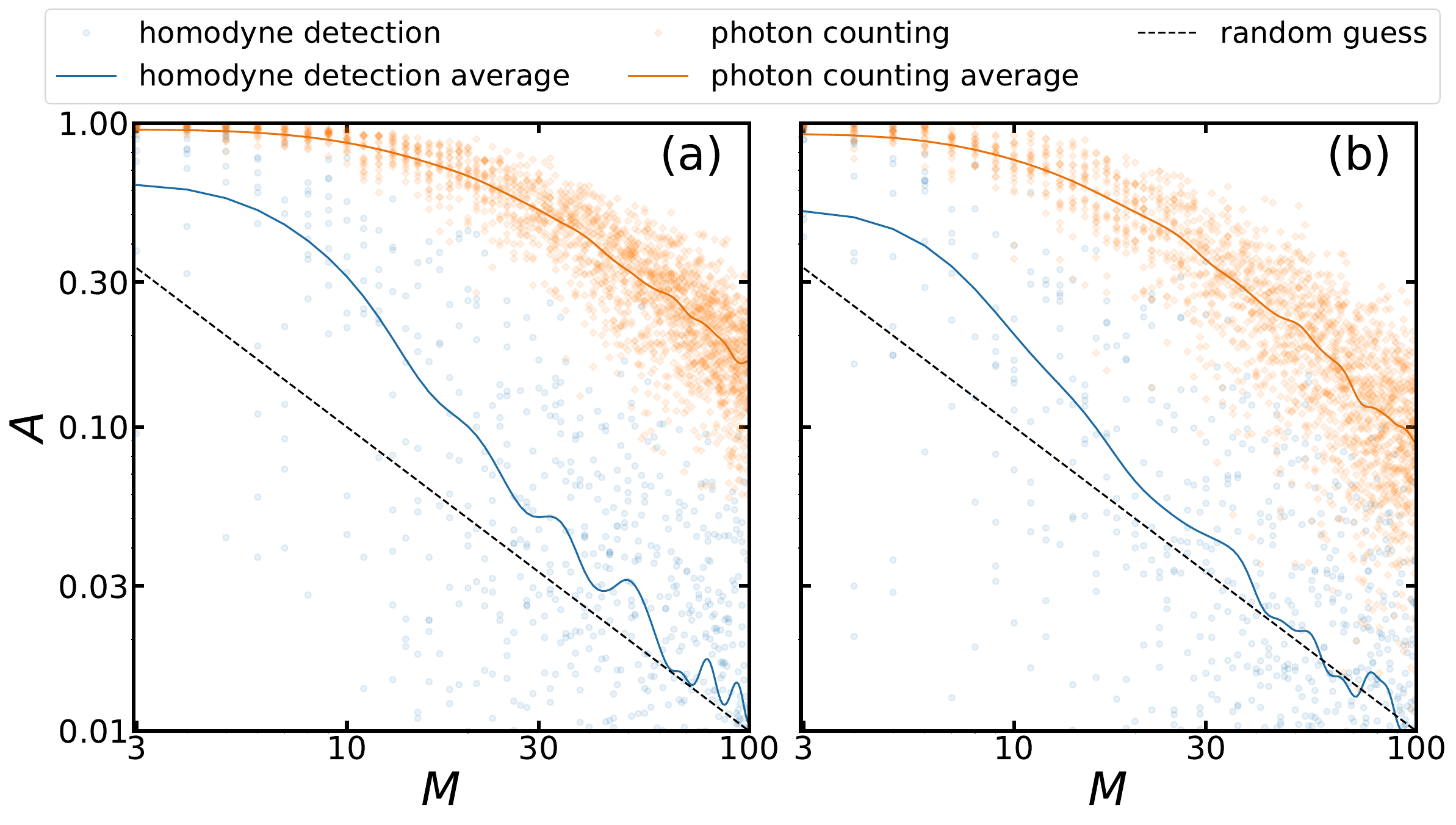}
    \caption{\label{fig:M} Accuracy as a function of the number of modes $M$ for (a) PCA and (b) CCA. Each data point corresponds to a separate training run with distinct initialization and signal sampling. We fix $\sqrt{2M} \sigma_c = 0.2$ to maintain constant total signal strength. Dashed lines indicate the expected random guess limit $1/M$.}
\end{figure}

We next study the effect of scaling the number of modes $M$, while holding the total signal strength fixed at \( \sqrt{M} \sigma_c = \text{const} \). In principle, this normalization preserves the signal-to-noise scaling of the problem, since we consider completely correlated noise (i.e., rank 1 covariance, $\bm V$, with eigenvalue $2M\sigma_c^2$). However, in practice, the optimization landscape becomes more complex as $M$ increases due to an increase in the number of parameters, and the training process may converge to suboptimal local minima. As shown in Fig.~\ref{fig:M}, the resulting accuracy decreases with increasing $M$, reflecting this growing optimization complexity.

For small values of $M$, both detection strategies perform well in terms of optimization complexity, albeit with photon counting still yielding much higher absolute accuracy. As $M$ increases, homodyne detection (blue) rapidly degrades toward the random guess limit, scaling approximately as $1/M$. Photon counting (orange), by contrast, maintains good performance across a broader range of $M$. At large $M$, photon counting has its scaling slowly moving towards the same $1/M$, while still maintains advantage over homodyne. In any case, photon counting consistently exhibits a significant performance advantage over homodyne detection across all values of $M$. This robustness indicates the resilience of photon counting to both quantum noise and training complexity in high-dimensional settings.

\edit{\section{Effects of Practical Imperfections}

In this section, we analyze departures from the idealized model, including photon-detector inefficiency and dark counts, dephasing in the linear-optical unitary, and non-Gaussian signals.

\subsection{Detector Efficiency and Dark Counts}

\begin{figure}
    \centering
    \includegraphics[clip = true, width =\linewidth]{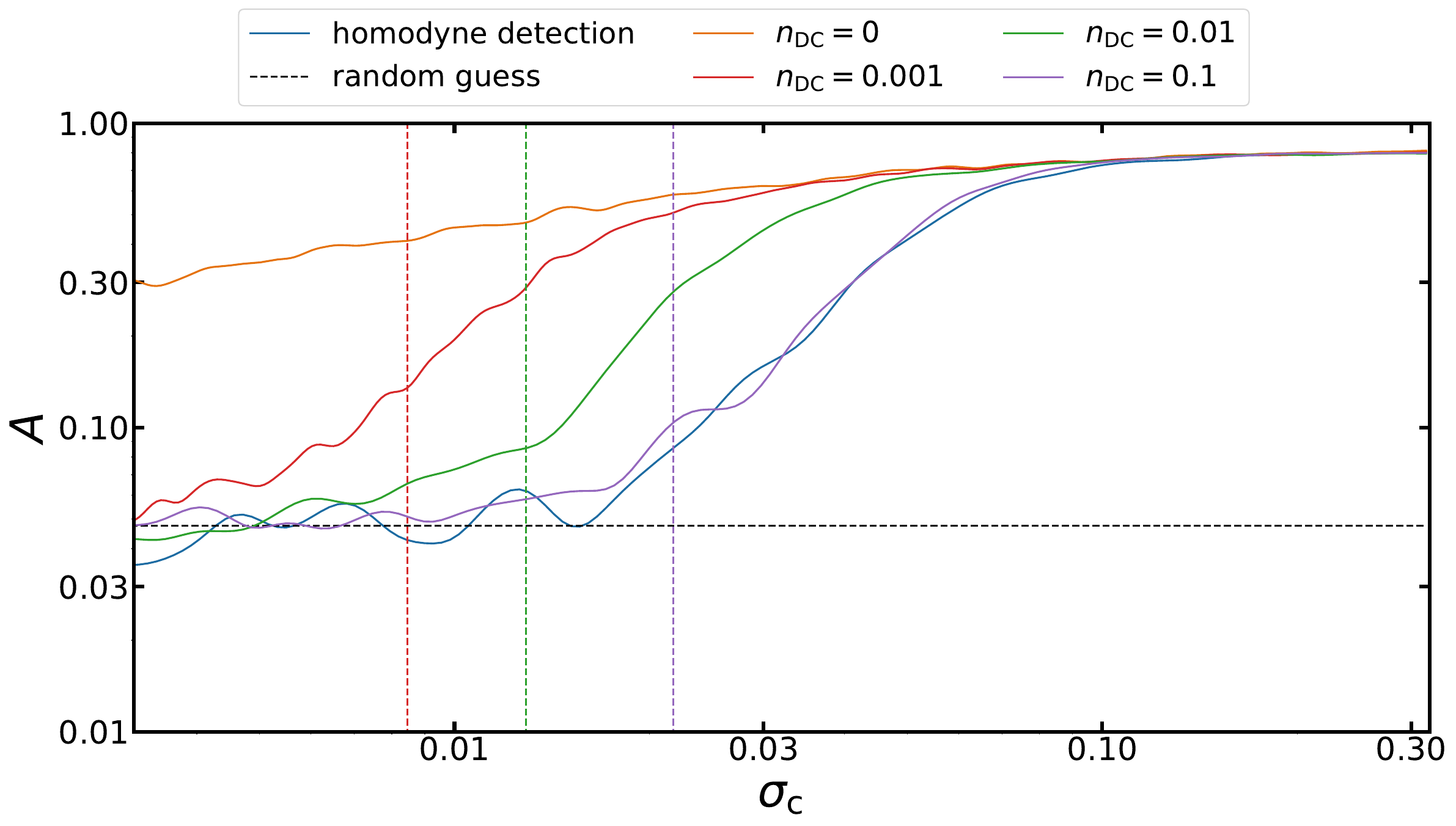}
    \caption{\label{fig:therm}  Average training performance of PCA in the presence of dark counts. Curves in different colors correspond to varying levels of effective thermal noise induced by dark counts ($n_{\rm DC}=\Gamma_{\rm DC}\tau$). The number of modes is fixed at $M=21$. Vertical dashed lines indicate the points where the signal-to-noise ratio (SNR) reaches 3\% under different thermal noise levels. }
\end{figure}

Photon detectors exhibit imperfections that significantly degrade estimation performance, particularly in the weak-signal regime. The primary culprits are finite detection efficiency and dark counts --- spurious detection events arising from background radiation or electronic instabilities. We assess the impact of such imperfections on PCA (similar analyses hold for CCA), presenting a simplified theoretical model alongside supporting numerical results.

Real photon detectors suffer from two primary imperfections: finite quantum efficiency and dark counts. Finite efficiency, denoted $\eta$, models the probability that an incident photon is successfully detected. Whereas dark counts arise from thermal or electronic noise and can be modeled as background excitations that appear in the absence of the signal.

We mathematically model an imperfect photon detector by a three-stage process: first, an attenuation channel with transmission $\eta$; second, a random displacement $\beta\sim\mathcal{CN}(0,n_{\rm DC})$ representing additive noise from dark counts, where $n_{\rm DC} = \Gamma_{\rm DC} \tau$ is the expected number of dark counts over an integration time $\tau$, $\mathcal{CN}$ is complex normal distribution; and third, ideal photon-number detection.

The input state to the imperfect photon detector is the output of the PCA variational processor, which is a single-mode, incoherently displaced state with displacement $\lambda_w\sim\mathcal{N}(0,\vartheta_{\rm PCA})$ along the $P$ quadrature [recall $\vartheta_{\rm PCA}=\vec{w}^T\bm V\vec{w}$]. After accounting for attenuation and dark counts, the effective state remains an incoherently displaced state, but now with variance $\eta\vartheta_{\rm PCA}+n_{\rm DC}/2$ along the $P$ quadrature and variance $n_{\rm DC}/2$ along the $Q$ quadrature due to dark counts affecting both quadratures equally. The average photon counts measured by the detector is thus
$\bar{n} = \eta\vartheta_{\rm PCA}/2 + n_{\rm DC}$. The first term originates from the signal and the second from the dark-count noise, which represents the detector noise floor. 

In the presence of detector inefficiency and dark counts, the loss observable must be augmented as
\begin{equation}
    \hat{L} = -\frac{2}{\eta}(\hat{N}_w - n_{\rm DC}),
\end{equation}
so that $\expval*{L}=-\vartheta_{\rm PCA}$, as before. In other words, the dark counts must be characterized and subtracted off, while the finite quantum efficiency requires rescaling of the estimator. Furthermore, the loss observable now suffers higher noise fluctuations. In the regime where dark counts dominate the total photon number, $\bar{n} \approx n_{\rm DC}$, we find that
\begin{equation}
    \mathrm{Var}(\hat{L}) \approx \frac{4}{\eta^2} n_{\rm DC}(n_{\rm DC} + 1).
\end{equation}
Comparing this to a perfect homodyne detector, for which $\eval*{\mathrm{Var}(\hat{L})}_{\rm hom} \approx 1/2$, we conclude that photon counting with an imperfect photon detector is advantageous when
\begin{equation}
    \frac{ n_{\rm DC}(n_{\rm DC} + 1)}{\eta^2} \ll \frac{1}{8}.
\end{equation}
For detectors with $\eta \gtrsim 0.9$ and dark-count rates $n_{\rm DC} \ll 1$, this condition is easily satisfied. Note that this analysis pertains to the physical estimation noise. Though how this translates to adaptively learning (e.g., learning the PCA vector $\vec{w}^\star$) is non-trivial. We provide quantitative numerical assessments to address this below.

Here we numerically investigate how dark counts affect photon counting performance. For simplicity, we assume perfect detection efficiency ($\eta=1$).

For the numerical analysis, we model the detector statistics as a Poisson process. In the absence of an input signal, the detector receives an average of $n_{\rm DC}$ spurious photons per unit time. Note that the signal (i.e., the output of the quantum processor) is a random coherent state with mean photon number per unit time $\bar{n}_{\rm sig}=\vartheta_{\rm PCA}/2$. So that, in the ideal case of no dark counts, the photon detection statistics follow a Poisson distribution with mean $\bar{n}_{\rm sig}$, i.e., ${\rm Poisson}(\bar{n}_{\rm sig})$. When dark counts are included, this distribution is effectively shifted to ${\rm Poisson}(\bar{n}_{\rm sig} + n_{\rm DC})$.

In Fig.~\ref{fig:therm}, we take PCA as an example to illustrate how performance degrades with the dark-count rate. Note that the average photon number $\bar{n}_{\rm sig}$ at the output is proportional to the raw signal $M \sigma_{\rm c}^2$.\footnote{Recall that $\bar{n}_{\rm sig}=\vartheta_{\rm PCA}/2=\vec{w}^T\bm V\vec{w}/2$ and that $\bm V=2M\sigma_c^2 \vec{v}\vec{v}^T$.} When $\bar{n}_{\rm sig} \gg n_{\rm DC}$, the effect of thermal noise is negligible, and the performance approaches that of the ideal detector. In contrast, when $\bar{n}_{\rm sig} \ll n_{\rm DC}$, noise dominates and photon-counting accuracy deteriorates sharply. 

We use the dashed line in Fig.~\ref{fig:therm} to indicate the signal strength at different dark-count levels, where the analytical calculations of the SNR coincide. The ratio between the signal amplitude and the noise amplitude is 3\%. The accuracy remains similar for different dark-count levels around 0.1, which marks the approximate lower bound of signal strength required to achieve a quantum advantage. Beyond this threshold, the accuracy improves significantly.

In practice, as long as the dark-count rate is sufficiently low such that $n_{\rm DC} < \bar{n}_{\rm sig}$, photon counting remains effective despite the presence of detector noise.

\subsection{Dephasing}

\begin{figure}
\centering
\includegraphics[clip = true, width =1\linewidth]{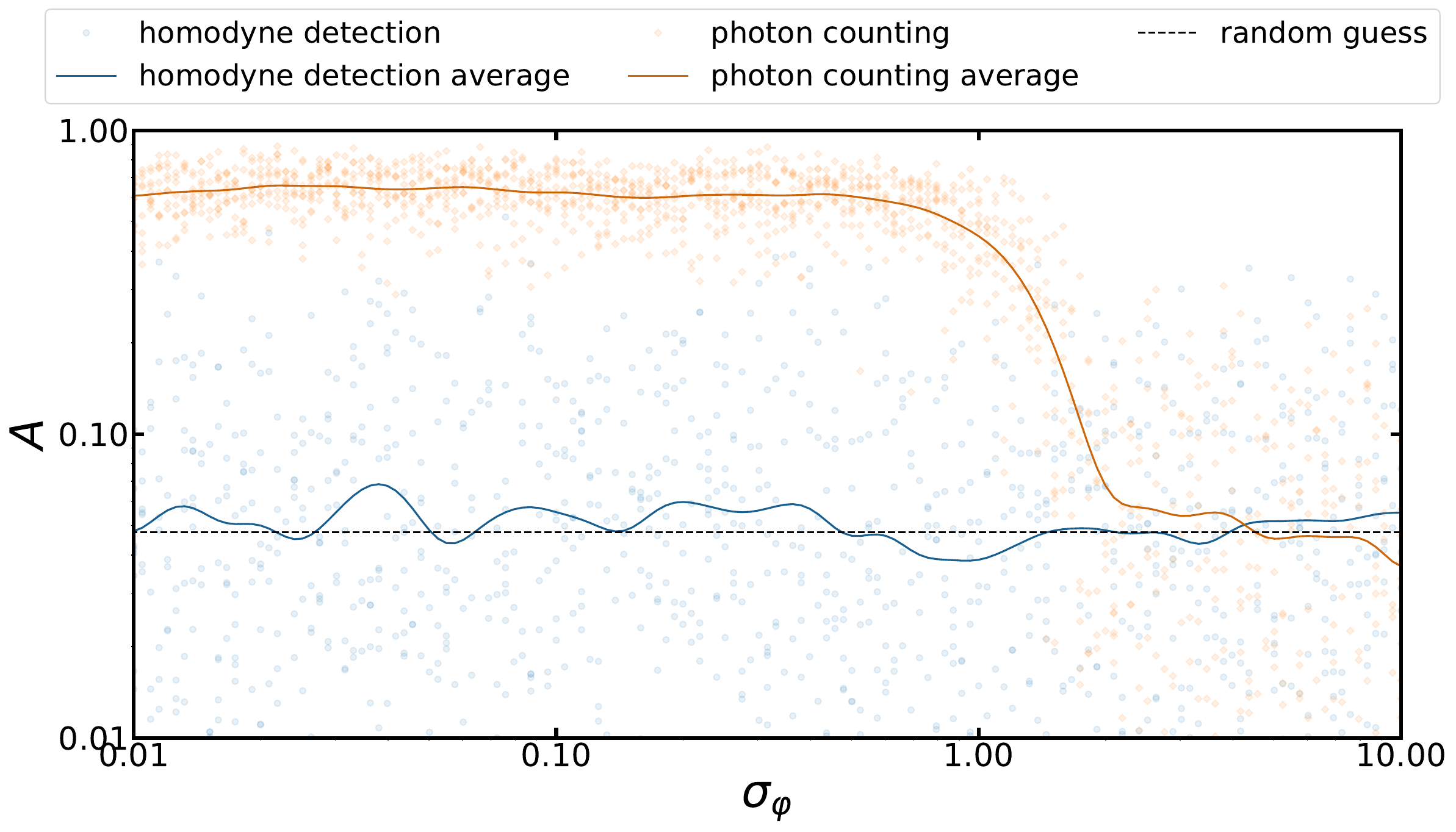}
\caption{\label{fig:dephase} Accuracy as a function of dephasing strength for PCA. The dephasing strength is $\sigma_\varphi=0.02$. Other parameters are $M=21$ and $\sigma_c=0.02$.}
\end{figure}

In practical settings, the physical setup is often susceptible to dephasing (i.e., random phase fluctuations), which ultimately degrades the performance in quantum sensing. It is more convenient, in this case, to describe the stochastic signal displacements as a complex vector $\vec{\lambda}_{\mathbb{C}}$ with $\Re\{\vec{\lambda}_{\mathbb{C}}\}=0$ and $\Im\{\vec{\lambda}_{\mathbb{C}}\}=i\vec{\lambda}/\sqrt{2}$. 
Dephasing is as a form of decoherence that (i) spreads the originally imaginary-only (momentum) displacement into both quadratures, thereby mixing in position components, and (ii) washes out correlations and coherences between displacements on different modes---the latter being the most damaging effect for interferometric sensing. We model this process as an independent random phase applied to each channel immediately after the signal imprinting but before any variational measurement or processing: $(\vec{\lambda}_{\mathbb{C}})_k \rightarrow (\vec{\lambda}_{\mathbb{C}})_k e^{i\varphi_k}$, where $\varphi_k$ is an iid random variable for each $k$. Variational processing and measurement is then performed on this dephased signal. 

In the case of photon counting, the dephased signal strength for PCA becomes
\begin{eqnarray}
    2N_w&=&\left|\sum_l  w_l \lambda_l\cos\varphi_l\right|^2+\left|\sum_l  w_l \lambda_l\sin\varphi_l\right|^2 \nonumber\\
    &=&\sum_l  \left|w_l \lambda_l\right|^2+\sum_{k \neq l}w_k \lambda_k w_l \lambda_l \cos(\varphi_k-\varphi_l) ~.
\label{eq:Ndephase}
\end{eqnarray}

For homodyne detection, only momentum quadrature is measured.
The corresponding signal strength of homodyne detection for PCA after dephasing becomes
\begin{eqnarray}
    P_w&=&\sum_l  w_l \lambda_l\cos\varphi_l \nonumber\\
    P_w^2&=&\left|\sum_l  w_l \lambda_l\cos\varphi_l\right|^2 ~.
\label{eq:Pdephase}
\end{eqnarray}

In the limit $\varphi_l\ll 1$, the signal strength is reduced by approximately $O(\varphi_l^2)$, indicating that our protocol is not overly sensitive to dephasing. Furthermore, since Eq. (\ref{eq:Ndephase}) exceeds Eq. (\ref{eq:Pdephase}), photon counting yields a stronger signal than homodyne detection. 

We assume that iid dephasing in each mode, with each following a Gaussian distribution $\varphi_l \sim \mathcal{N}(0,\sigma_{\varphi})$. In Fig. \ref{fig:dephase}, we illustrate the weak-signal case for PCA. The results show that photon counting remains robust against weak dephasing, as indicated by the extended plateau of the curve. A noticeable drop in accuracy occurs only when the dephasing becomes very strong.

\subsection{Non-Guassian Signals}
In our benchmark problems, the signals are modeled as Gaussian displacements, where the covariance matrix fully characterizes the data. This assumption, however, neglects the possibility of non-Gaussian statistics, which are often relevant in practical sensing scenarios. To illustrate this, we consider a bimodal Gaussian distribution, $\lambda \sim \left(\mathcal{N}(-\sqrt{10M}\sigma_c, \sqrt{2M}\sigma_c)+\mathcal{N}(\sqrt{10M}\sigma_c, \sqrt{2M}\sigma_c)\right)/2$. As shown in Fig. \ref{fig:DuoGua}, our results demonstrate that the proposed method continues to perform effectively even in the presence of such non-Gaussian signals.

\begin{figure}
    \centering
    \includegraphics[clip = true, width =1\linewidth]{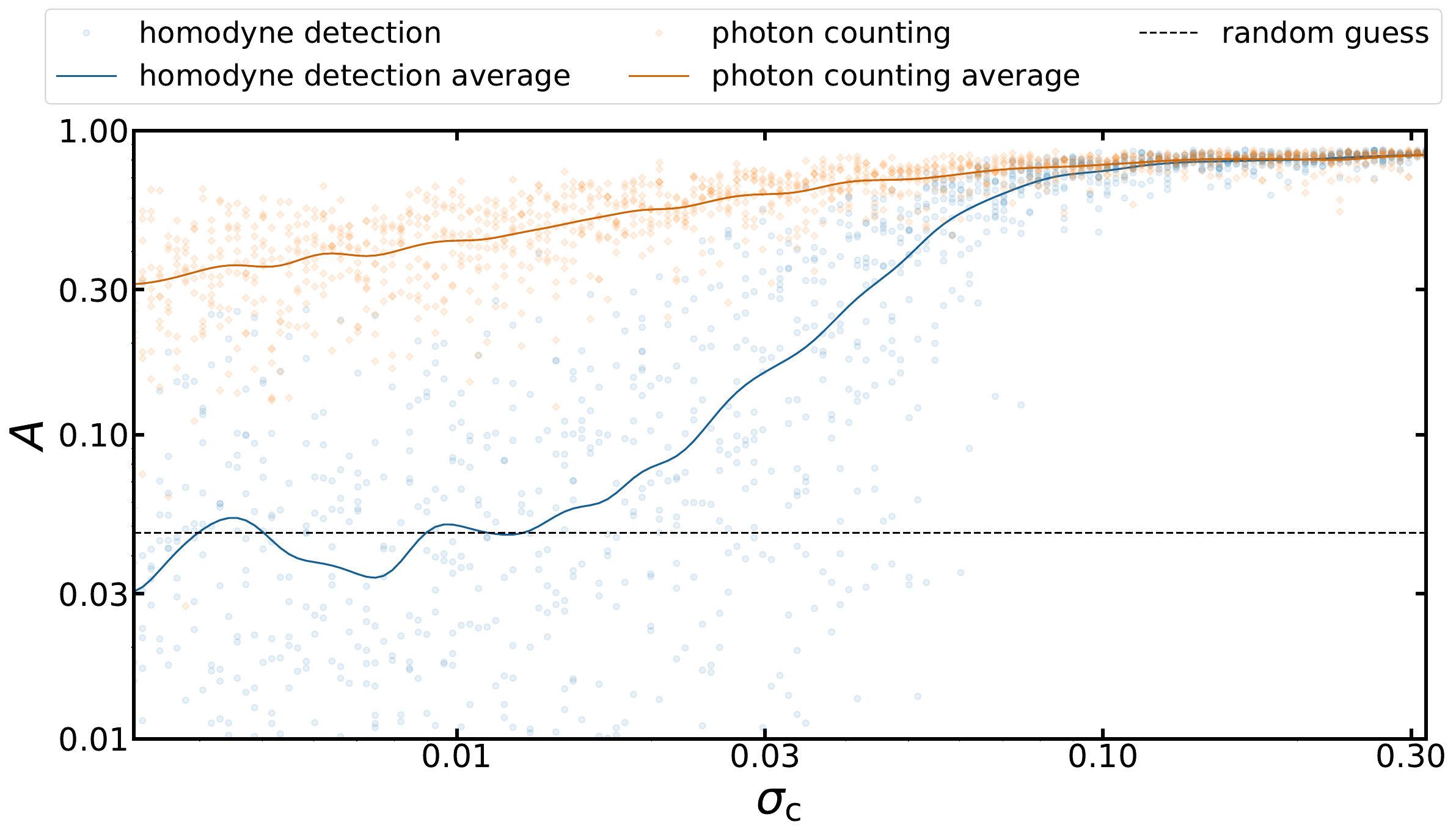}
    \caption{\label{fig:DuoGua} Accuracy versus non-Gaussian signal strength for PCA. Other parameters are $M=21$ and $\sigma_c=0.02$.}
\end{figure}
}


\section{Further enhancement from squeezing}
\label{sec:squeezing}

\begin{figure}
    \centering
    \includegraphics[width=\linewidth]{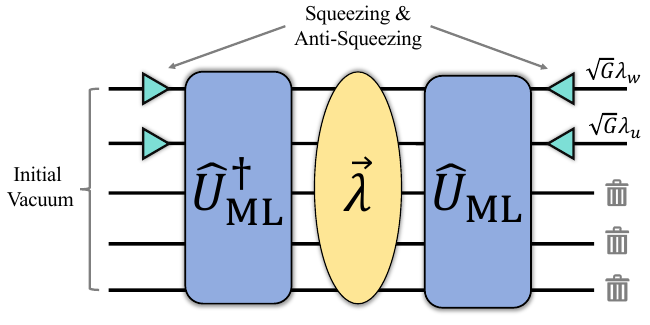}
    \caption{Distributed squeezed-vacua assisted variational quantum processor. Stochastic displacements are encoded onto quantum probes initialized in a continuous-variable entangled state in the form of distributed squeezed vacua~\cite{Zhuang2018PRA_dqs,Zhang2021DQSrvw}. Anti-squeezing at the receiving end coherently amplifies the weak stochastic signals relative to vacuum fluctuations, permitting a quantum enhancement in variational parameter estimation and associated learning tasks.}
    \label{fig:squeezed}
\end{figure}

The accuracy of our protocols can be enhanced by leveraging squeezing in tandem with variational quantum processing, as illustrated in Fig.~\ref{fig:squeezed}. The improvement originates from the use of a continuous-variable entangled probe state~\cite{Zhuang2018PRA_dqs,Zhuang2019SLAEN_theory,Zhang2021DQSrvw} and by performing joint quantum measurements across the distributed modes~\cite{Brady2024CorrNoise}. 

A simpler (though more mundane) way to understand the benefit of squeezing, without referring to entanglement, is as follows: If we lump all pre- and post-quantum processing together, including all squeezing and variational operations, then the net transformation equates to simply amplifying the stochastic displacements without adding extra (thermal nor vacuum) noise, such that the effective transform is simply  $(\lambda_w,\lambda_u)\rightarrow\sqrt{G}(\lambda_w,\lambda_u)$. Accordingly, the second moments transform as ${\mathbb{E}[\lambda_i\lambda_j]\rightarrow G\times\mathbb{E}[\lambda_i\lambda_j]}$ with $i,j\in\{w,v\}$. This transformation implies that the variational estimation tasks remain formally identical to the unsqueezed case, but with the input signals now coherently amplified. Hence, we can infer the performance accuracy of the distributed squeezed-vacua assisted protocol directly from prior data (Fig.~\ref{fig:CN}) by simply rescaling the input signal strength to $\sqrt{G}\sigma_c$ and referencing the corresponding accuracy. We emphasize that phase-sensitive (i.e., coherent) amplification is paramount here. Contrast this with phase-insensitive amplification, which would inevitably add vacuum noise to the output~\cite{Caves1982noise}, thereby drowning any potential advantage afforded by subsequent photon counting measurements. 

The distributed squeezed-vacua assisted variational quantum processor has a similar flavor to the original SLAEN recipe~\cite{Zhuang2019SLAEN_theory} but, crucially, we augment the detection stage to include anti-squeezing operations and photon counting measurements (see also Refs.~\cite{Gorecki2022SpreadChannel,Tsang2023NoiseSpectr} for its relevance in single-mode parameter estimation). This modification at the receiver end enhances sensitivity to weak stochastic signals, surpassing the performance achievable with homodyne detection considered in Refs.~\cite{Zhuang2019SLAEN_theory,Xia2021PRX_SLAEN}. The induced advantage of the proposed scheme over the original SLAEN is identical to the advantage of photon counting over homodyne, due to the amplification of signals applying to both current protocol and original SLAEN. \edit{Including variational quantum processing at the front-end is closer in spirit to a full variational quantum toolbox~\cite{Meyer2021:varToolbox}.}

\section{Outlook}

Our results show that photon counting offers a clear advantage over homodyne detection for high-dimensional physical-layer learning tasks in the weak-signal regime, where vacuum fluctuations limit homodyne performance. In both principal component analysis (PCA) and cross-correlation analysis (CCA), photon counting consistently achieved higher accuracy, even under training noise and sampling variability. These findings highlight the robustness of photon counting as a strategy for high-precision quantum signal processing and learning of weak, high-dimensional, physical-layer data.

The proposed scheme bears conceptual similarity to the SLAEN framework~\cite{Zhuang2019SLAEN_theory,Xia2021PRX_SLAEN}, but crucially differs in its detection strategy. Whereas SLAEN employed homodyne measurements with classical post-processing, our approach incorporates programmable linear-optics directly at the measurement stage, followed by photon counting. This tradeoff --- replacing classical post-processing with physical-layer variational quantum processing and homodyne detection with photon counting --- yields a substantial gain in sensitivity and performance. \edit{This approach is also reminiscent of structured receivers developed for coherent-state discrimination~\cite{Guha2011StructuredRec, Sidhu2021PSK, Sidhu2023Discrim}.}

Beyond this baseline architecture, we explored enhancements from distributed squeezed-vacua. The resulting squeezed-vacua assisted variational processor --- featuring a sandwich structure of squeezing, trainable linear-optical transformations, and anti-squeezing --- enables coherent amplification of the signal prior to photon detection. This strategy translates into a direct scaling benefit for variational estimation tasks. That said, the presence of excitation loss (not treated here) imposes practical limitations on this strategy, motivating future extensions using idler-based entanglement (viz., two-mode squeezing with quantum memory), as analyzed in Refs.~\cite{Shi2023DMLimits,Shi2025rayleigh,Gardner2024StochEst} for a single mode.

The methods developed here are well-suited to emerging application-specific domains in quantum-enhanced estimation and learning. These include distributed quantum sensing networks for fundamental physics applications~\cite{Ahmed2018QuSenseHEP, Bass2024NatRvw, YeZoller2024PRL_Essay}, such as searching for dark matter~\cite{HAYSTAC2021QuDMSearch,Brady2022QuNetworkDMSearch,Brady2023OmechArray,Agrawal2024FockRadiometer} and stochastic gravitational wave signatures~\cite{Mcculler2022photoncount,Vermeulen2025gquest} \edit{(see Section~\ref{sec:phys-relev} for further discussion)}. 

Recent efforts in bosonic quantum learning~\cite{Oh2024LearnOscillators,Liu2025bosonicLearning} have demonstrated that entanglement-assisted strategies based on two-mode squeezing can significantly reduce the sample complexity of learning multimode displacement channels. In high-dimensional settings, our variational interferometric photon-counting module could be used as a front-end to compress (weak) input data onto a small number of relevant modes via physical-layer dimensionality reduction (e.g., through PCA). This would be used to compress high-dimensional multimode displacement channels into a smaller set of modes, to which a smaller number of two-mode squeezers are applied, thus reducing the physical resource cost while retaining sensitivity to the relevant signal subspace. 

\edit{Quantum sensing with a variational network of mechanical oscillators, such as trapped ions~\cite{Gilmore2021IonEFieldQSN} or optomechanical systems~\cite{Xia202PRL_RadioQSN, Brady2023OmechArray, Xia2023OmechDQS}, provides a natural playground for our methods. Stochastic forces or fields with rich spatial distributions can drive the sensor array, and identifying the most informative collective modes is paramount. Assessing correlations between collective modes in the array may be interesting from a physics viewpoint.}

Our variational approach may also prove useful in the domain of sub-diffraction incoherent imaging~\cite{Tsang2016Superresolution,Lupo2020PRL_QuLinearLimits, Grace2022Imaging, Buonaiuto2025ML_subdiff}, where optimal performance is known to rely on linear optics and photon counting. In this context, learning the first few principle components of a complex multi-emitter scene through variational quantum measurements may provide a practical path towards dispelling ``Rayleigh's curse"~\cite{Tsang2019Starlight}. 



\section*{acknowledgements}
QZ acknowledges discussions with Zheshen Zhang. This work is supported by Office of Naval Research Grant No. N00014-23-1-2296. QZ also acknowledges support from National Science Foundation Grant No. 2350153 and Defense Advanced Research Projects Agency (DARPA) D24AC00153-02.

\section*{Data availability}
The codes and data that support the findings of this article are
available at \cite{github}.

%

\end{document}